\documentclass[doublecol]{epl2}

\title{Special relativity with an arbitrary invariant limiting velocity of a particle}
\shorttitle{Special relativity with an arbitrary invariant limiting velocity of a particle}

\author{A. S. Parvan\inst{1,2}}
\shortauthor{A. S. Parvan }

\institute{
  \inst{1} Bogoliubov Laboratory of Theoretical Physics, Joint Institute for Nuclear Research, 141980 Dubna, Russian Federation\\
  \inst{2} Institute of Applied Physics, Moldova Academy of Sciences, MD-2028 Chisinau, Republic of Moldova
}


\abstract{It is shown that a generalized special theory of relativity (GSTR) with an arbitrary limiting velocity of a particle different or equal to the speed of light in vacuum can be constructed from the canonical equation of the $4$-dimensional hyperboloid of revolution. In particular, when the limiting velocity equals the speed of light, the special theory of relativity (STR), which corresponds to the equation of the equilateral hyperboloid of revolution, is recovered. The generalized Lorentz transformations for any values of the speed limit for both the time-like and the spice-like regions in the coordinate and momentum spaces and in the general parametric form were obtained. It was established that the inversion of time (energy) axis under the (generalized) Lorentz transformations is forbidden. In the GSTR, the rest mass of a space-like particle is real. Our results strongly suggest that the muon neutrino in the OPERA experiment might most likely be a time-like or a light-like superluminal particle, whose limiting velocity may exceed the speed of light in vacuum, rather than a superluminal space-like particle (tachyon) with a speed limit equal to speed of light for which the rest mass $m=117.1^{+11.0}_{-10.5}$ MeV$/c^{2}$.}

\begin{document}

\maketitle

\section{Introduction}
Particles propagating with velocities that are greater than the velocity of light in vacuum are precluded by the validity of the special theory of relativity and the principle of causality~\cite{Einstein1,Pauli,Landau,Brillouin}. The limiting velocity of any particle is considered to be equal to the velocity of light (electromagnetic wave) in vacuum. This was accepted on the basis of the invariance of the speed of light and of Maxwell's equations in all inertial reference frames. Yet, there is no experimental confirmation that no wave-particle can move faster than the speed of light. Therefore, there exists an experimental search for the signals that propagate with velocities that are greater than the velocity of light in vacuum~\cite{Alvager,Mugnai,Wang,Carey}. Moreover, the OPERA Collaboration has recently reported the observation of superluminal muon neutrinos~\cite{Opera}, whose speed $v$ exceeds that of light $c$, with $(v-c)/c=[2.37\pm 0.32(stat)^{+0.34}_{-0.24}(sys)]\times 10^{-5}$. In theoretical attempt to define the particles of super-luminary velocities, which were initially elaborated by Sommerfeld~\cite{Sommerfeld}, several different possibilities exist. The simplest two examples of these, which are considered in the framework of STR, consist in introducing the space-like particles (tachyons)~\cite{Bilaniuk,Feinberg,Recami} whose four-momenta are always space-like and whose velocities are therefore always greater than $c$ and in deforming the Minkowski space-time~\cite{Cardone}. However, several mathematical deficiencies in the definition of STR in the space-like region result in physical inconsistencies for the possibility of faster-than-light particles~\cite{Feinberg,Recami}. For instance, the space-like and the time-like regions embody the same invariant relation of the four-momentum and mass without taking into account the origin of the definition of the square length of a four-vector in STR. This leads to the imaginary rest mass of the space-like particles. Other inconsistencies arise from the application of the same Lorentz transformations to both the space-like and the time-like regions~\cite{Feinberg}.

\begin{figure*}
\onefigure{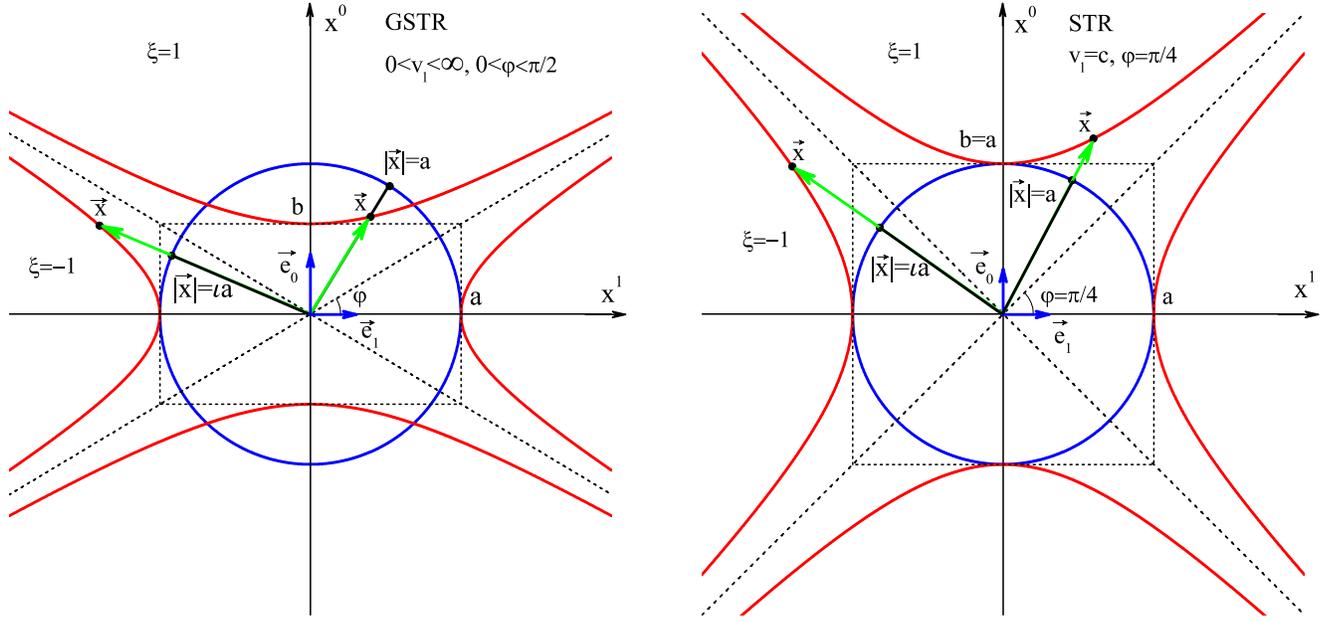}
\caption{(Color online) The magnitude $|\vec{x}|=\sqrt{(x^{0}/\tan\varphi)^{2}-(x^{1})^{2}-(x^{2})^{2}-(x^{3})^{2}}=\sqrt{\xi}a$ of a four-vector $\vec{x} = x^{i} \vec{e}_{i}$ in the GSTR (left panel) and the STR (right panel) at $x^{2}=x^{3}=0$.}
\label{fig.1}
\end{figure*}
The main purpose of this Letter is to show on the basis of the tensor analysis that the (STR) GSTR can be completely deduced in a unified manner from the canonical equation of the $4$-dimensional (equilateral) hyperboloid of revolution and is to apply this generalized theory to the study of the OPERA superluminal neutrino anomaly at LHC. The GSTR is obtained from the STR by replacing a particular equilateral hyperboloid of revolution with a hyperboloid of revolution defined in the general form. Such a unified definition of the scalar product and metrics of the (STR) GSTR from the (equilateral) hyperboloid of revolution allows one to solve the problem of imaginary mass for superluminal space-like particles (tachyons) and, moreover, permits one to introduce any values of limiting velocity, which results in possible existence of three different types of superluminal particles. The (generalized) Lorentz transformations for any values of the speed limit for both the time-like and the spice-like regions in the coordinate and momentum spaces and in the general parametric form are found.

\section{GSTR formulation}
Let the contravariant coordinates $x^{i} = \vec{x} \vec{e}^{i}$ of the four-vector $\vec{x} = x^{i} \vec{e}_{i}$ in the $4$-dimensional linear vector space $L$, which is defined by the basis $\{\vec{e}_{i}\}$ $(i=0,1,2,3)$~\cite{Pauli,Charlier}, satisfy the canonical equation for the $4$-dimensional hyperboloid of revolution
\begin{equation}\label{1}
    \frac{(x^{0})^{2}}{b^{2}} - \frac{(x^{1})^{2}}{a^{2}} - \frac{(x^{2})^{2}}{a^{2}}  - \frac{(x^{3})^{2}}{a^{2}} = \xi, \quad
    \tan \varphi = \frac{b}{a},
\end{equation}
where $a$ and $b$ are the space and time semi-axes of the hyperboloid, $0<\varphi<\pi/2$ is the angle between the asymptotes and the space axes $x^{1},x^{2}$ and $x^{3}$, the parameter $\xi=1$ in the time-like region at $|x^{0}| > [(x^{1})^{2}+(x^{2})^{2}+(x^{3})^{2}]^{1/2} \tan \varphi$, $\xi=-1$ in the space-like region at $|x^{0}|< [(x^{1})^{2}+(x^{2})^{2}+(x^{3})^{2}]^{1/2} \tan \varphi$ and $\xi=0$ in the light-like region at $|x^{0}|= [(x^{1})^{2}+(x^{2})^{2}+(x^{3})^{2}]^{1/2} \tan \varphi$. Here and in the following we consider a summation on each repeated index. Let the square length $\vec{x}^{2}=g_{ij} x^{i} x^{j}$ of a $4$-vector $\vec{x}$, where $g_{ij}=\vec{e}_{i}\vec{e}_{j}$ is a metric tensor~\cite{Pauli}, be defined by the product of a square length of a semi-axis $a$ and the parameter $\xi$ of the hyperboloid (\ref{1}) (see Fig.~\ref{fig.1}),
\begin{equation}\label{2}
    \vec{x}^{2} = \xi a^{2} = \frac{1}{\tan^{2}\varphi}(x^{0})^{2}-(x^{1})^{2}-(x^{2})^{2}-(x^{3})^{2}.
\end{equation}
Then, the components of $g_{ij}$ and the inverse metric tensor $g^{ij}=\vec{e}^{i}\vec{e}^{j}$, which is defined by the reciprocal basis $\{\vec{e}^{i}\}$ $(i=0,1,2,3)$ and satisfies the unitary equation $g^{ij}g_{jk}=\delta^{i}_{\ k}$~\cite{Pauli,Charlier}, can be written as
\begin{eqnarray}\label{3}
   g_{ij}&=&(\frac{1}{\tan^{2}\varphi}, -1, -1, -1)  \\ \label{4}
   g^{ij}&=&(\tan^{2}\varphi, -1, -1, -1),
\end{eqnarray}
$g_{ij}=g^{ij}=0$ for $i\neq j$. The covariant coordinates $x_{i} = \vec{x} \vec{e}_{i}=g_{ij} x^{j}$ of the four-vector $ \vec{x} = x_{i} \vec{e}^{i}$ for which $x^{i}=g^{ij} x_{j}$~\cite{Pauli,Charlier} are written as
\begin{equation}\label{5}
    x_{0}=\frac{x^{0}}{\tan^{2}\varphi}, \; x_{1}=-x^{1}, \; x_{2}=-x^{2}, \; x_{3}=-x^{3}.
\end{equation}
Using the definitions of the metric and inverse metric tensors, the definition of reciprocal basis $ \vec{e}^{i}\vec{e}_{j} = \delta^{i}_{\ j}$ and Eqs.~(\ref{3}), (\ref{4}), we explicitly obtain  $\vec{e}_{0}  = (1/\tan\varphi,0,0,0)$, $\vec{e}_{1}=(0,\imath,0,0)$, $\vec{e}_{2} = (0,0,\imath,0)$, $\vec{e}_{3}=(0,0,0,\imath)$ and $\vec{e}^{0}  = (\tan\varphi,0,0,0)$, $\vec{e}^{1}=(0,-\imath,0,0)$, $\vec{e}^{2} = (0,0,-\imath,0)$,  $\vec{e}^{3}=(0,0,0,-\imath)$. Thus, we have found that the spacial basis vectors $\vec{e}_{1},\vec{e}_{2},\vec{e}_{3}$ and $\vec{e}^{1},\vec{e}^{2},\vec{e}^{3}$ take imaginary values. Therefore, the spacial subspace of the Minkowski space is complex. Note that the STR corresponds to $\varphi=\pi/4$.

In {\it the coordinate and momentum spaces} we have $x^{0}=ct,x^{1}=x,x^{2}=y,x^{3}=z$, $b=cT_{0},a=\lambda_{0}$ and $x^{0}=E/c,x^{1}=p_{x},x^{2}=p_{y},x^{3}=p_{z}$, $b=E_{0}/c$, $a=\mathcal{P}_{0}$, respectively, where $t$ and $E$ are the time and the energy, $x,y,z$ are the spacial coordinates, $p_{x},p_{y},p_{z}$ are components of the $3$-momentum~\cite{Landau}, $T_{0}$ and $\lambda_{0}$ are some parameters, $E_{0}$ is the rest energy, and $\mathcal{P}_{0}$ is the minimal momentum of the particle at energy $E=0$. Using Eq.~(\ref{1}) we obtain
\begin{equation}\label{6}
 \tan\varphi =\frac{c}{v_{l}} =\frac{E_{0}}{c \mathcal{P}_{0}},
\end{equation}
where $v_{l} = \lambda_{0}/T_{0}$ is the speed limit of the particle, $0<v_{l}<\infty$. Adopting Einstein's famous formula for the mass-energy equivalence, $E_{0}= mc^{2}$, where $m$ is the mass, and using Eq.~(\ref{6}), we get $\mathcal{P}_{0}=mv_{l}$.

The square length of the $4$-vector (\ref{2}), $\vec{x}^{2}= g_{ij}x^{i}x^{j}=g^{ij}x_{i}x_{j}$, and the metrics (\ref{3}), (\ref{4}) in the coordinate and momentum spaces can be rewritten as
\begin{eqnarray}\label{7}
    \vec{s}^{2} &=& \xi \lambda_{0}^{2} = v_{l}^{2}t^{2}-x^{2}-y^{2}-z^{2}, \\  \label{8}
     \vec{p}^{2} &=& \xi m^{2} v_{l}^{2} = \frac{v_{l}^{2}}{c^{4}}E^{2}-p_{x}^{2}-p_{y}^{2}-p_{z}^{2},
\end{eqnarray}
and
\begin{equation}\label{9}
 g_{ij}=(\frac{v_{l}^{2}}{c^{2}}, -1, -1, -1), \;  g^{ij}=(\frac{c^{2}}{v_{l}^{2}}, -1, -1, -1),
\end{equation}
$g_{ij}=g^{ij}=0$ for $i\neq j$, where $\vec{s}$ is the four-interval, $\vec{p}$ is the four-momentum, $\xi=1$ for the time-like particles $(\vec{s}^{2}>0,\vec{p}^{2}>0)$, $\xi=-1$ for the space-like particles $(\vec{s}^{2}<0,\vec{p}^{2}<0)$ and $\xi=0$ for the light-like particles $(\vec{s}^{2}=0,\vec{p}^{2}=0)$. Then, the covariant components (\ref{5}) in the coordinate and momentum spaces are $x_{0}=v_{l}^{2}t/c$, $x_{1}=-x$, $x_{2}=-y$, $x_{3}=-z$ and $x_{0}=(v_{l}^{2}/c^{3})E$, $x_{1}=-p_{x}$, $x_{2}=-p_{y}$, $x_{3}=-p_{z}$, respectively, and  the time basis vectors are $\vec{e}_{0}=(v_{l}/c,0,0,0)$ and $\vec{e}^{0}=(c/v_{l},0,0,0)$. Thus, the rest mass $m$ of a particle for the space-like four-momentum $\vec{p}^{2}< 0$ is real.

For a free particle, if we write the relation $x^{0}/x^{\alpha}=const$, ($\alpha=1,2,3$), in the coordinate and momentum spaces, we get~\cite{Landau}
\begin{equation}\label{10}
    \frac{p_{x}}{E}= \frac{v_{x}}{c^{2}}, \qquad   \frac{p_{y}}{E}= \frac{v_{y}}{c^{2}}, \qquad  \frac{p_{z}}{E}= \frac{v_{z}}{c^{2}},
\end{equation}
where $v_{x}=x/t$, $v_{y}=y/t$ and $v_{z}=z/t$ are the components of a $3$-velocity. Using Eqs.~(\ref{8}), (\ref{10}), we obtain
\begin{eqnarray}\label{11}
  E &=& \frac{c^{2}}{v_{l}} \sqrt{p_{x}^{2}+p_{y}^{2}+p_{z}^{2} + \xi m^{2} v_{l}^{2}}, \\ \label{12}
  E &=& \gamma m c^{2},  \\ \label{13}
  p_{x} &=&  \gamma m v_{x}, \quad  p_{y} =  \gamma m v_{y}, \quad  p_{z} =  \gamma m v_{z},  \\ \label{14}
 \gamma &=& \frac{1}{\sqrt{\frac{1}{\xi}(1-\frac{v^{2}}{v_{l}^{2}})}},
\end{eqnarray}
where $v^{2}=v_{x}^{2}+v_{y}^{2}+v_{z}^{2}$, $\xi=1$ for $v<v_{l}$, $\xi=-1$ for $v>v_{l}$ and $\xi=0$ for $v=v_{l}$. The light-like particle is a massless  particle, $m=0$, with the velocity of motion $v=v_{l}$. However, the time-like and space-like particles are massive and their velocities are $v<v_{l}$ and $v>v_{l}$, respectively. They are separated by a wall of infinite energy. Hence as $v\to c$ from above or below, the energy and momentum become infinite. At energy $E=0$ (vacuum), the space-like particle moves with the velocity $v=\infty$ and has the minimal momentum $\mathcal{P}_{0}$~\cite{Recami}.

\begin{figure}
\onefigure[width=8.5cm]{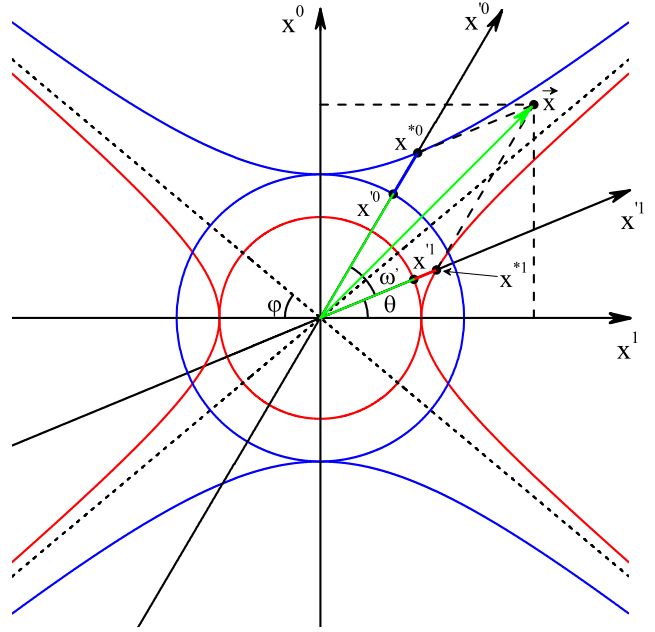}
\caption{(Color online) The generalized Lorentz transformations for $0<\varphi<\pi/2$. The Lorentz transformations correspond to $\varphi=\pi/4$.}
\label{fig.2}
\end{figure}

\section{Generalized Lorentz transformations}
\begin{figure}
\onefigure[width=8.5cm]{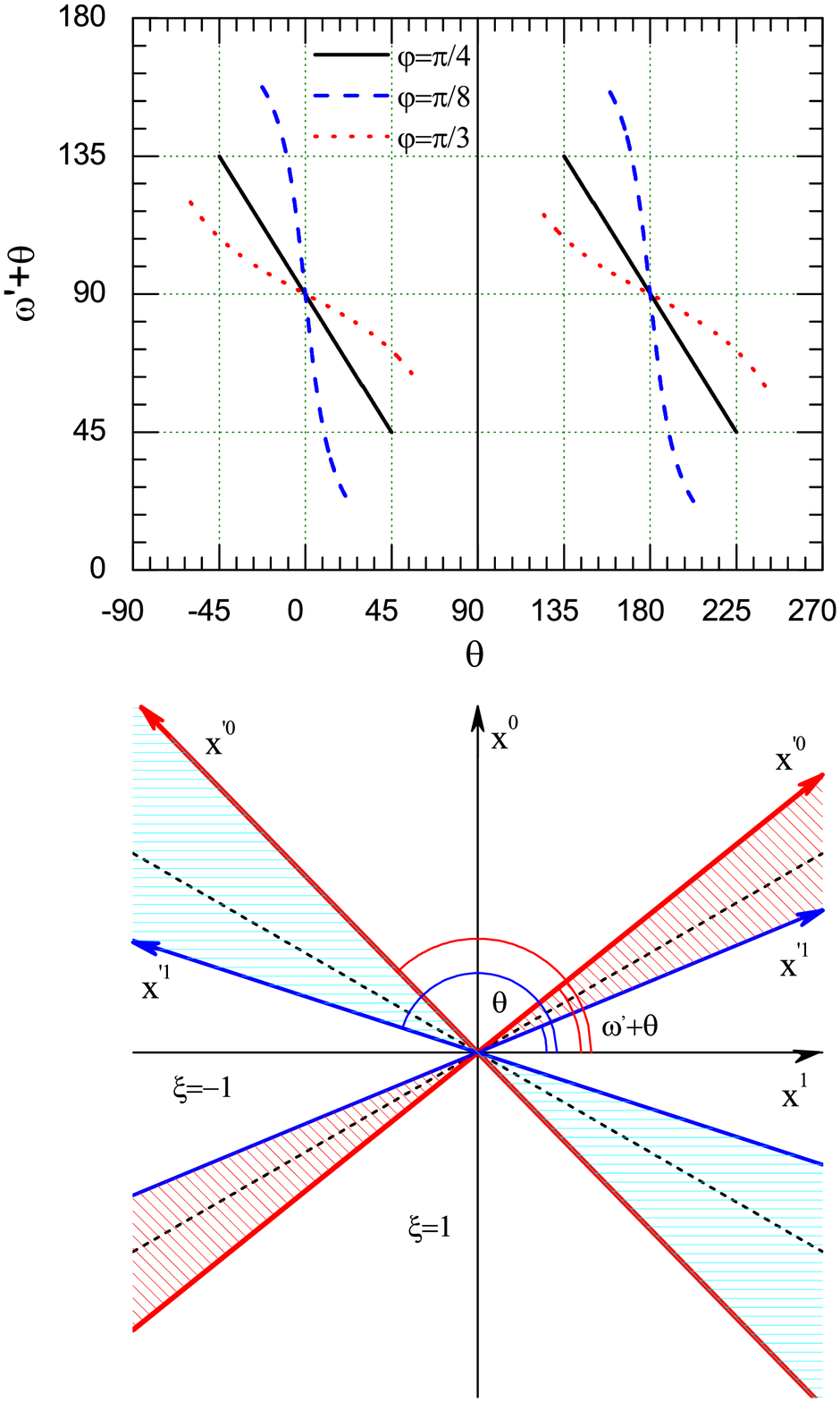}
\caption{(Color online) The dependence of angle $\omega'+\theta$ on the angle $\theta$ at different values of $\varphi$ for the GSTR. The angles are constrained by the relation $\cot(\omega'+\theta)=\tan \theta/\tan^2 \varphi$. The STR corresponds to $\varphi=\pi/4$.}
\label{fig.3}
\end{figure}

The generalized Lorentz transformation $x^{i} = a^{i}_{\ j} x'^{j}$, where $a^{i}_{\ j}=\partial x^{i}/\partial x'^{j}$ is the Jacobian matrix, can be written as the product of two transformations $a^{i}_{\ j}=(\Lambda_{O})^{i}_{\ k}(\Lambda_{D})^{k}_{\ j}$ in the plane $x^{0}-x^{1}$, i.e., the rotations of the oblique coordinate system $x^{i} = (\Lambda_{O})^{i}_{\ k} x^{*k}$ and the hyperbolic dilation of its axes (homothety) $x^{*k} = (\Lambda_{D})^{k}_{\ j} x'^{j}$, which is a projection of the hyperbola on the circle (see Fig.~\ref{fig.2}). We have
\begin{eqnarray}\label{15}
 a^{i}_{\ j} &=& \left(
               \begin{array}{cccc}
                 \eta_{0}\sin(\omega'+\theta) & \eta_{1}\sin\theta & 0 & 0 \\
                 \eta_{0}\cos(\omega'+\theta) & \eta_{1}\cos\theta & 0 & 0 \\
                          0           &      0     & 1 & 0 \\
                          0           &      0     & 0 & 1 \\
               \end{array}
             \right), \\ \label{16}
  \eta_{0} &=& \frac{1}{\sqrt{\frac{1}{\xi_{0}}[\sin^{2}(\omega'+\theta)-\tan^{2}\varphi \cos^{2}(\omega'+\theta)]}}, \;\; \\ \label{17}
  \eta_{1} &=& \frac{1}{\sqrt{\frac{1}{\xi_{1}}[\frac{\sin^{2}\theta}{\tan^{2}\varphi}- \cos^{2}\theta]}},
\end{eqnarray}
where $\omega'$ is the angle between the axes $x'^{0}$ and $x'^{1}$, $\theta$ is the angle between the axes $x'^{1}$ and $x^{1}$, and $\eta_{0}$, $\eta_{1}$ are the hyperbolic dilation parameters which are diagonal elements of the matrix $(\Lambda_{D})^{i}_{\ j}=(\eta_{0},\eta_{1},1,1)$, $(\Lambda_{D})^{i}_{\ j}=0$ for $i\neq j$. From the condition $g'_{ij}=a_{i}^{\ k} g_{kl} a^{l}_{\ j} = g_{ij}$, which preserves the length of the four-vector in the form $\vec{x}^{2} = g_{ij} x^{i} x^{j} = g_{ij} x'^{i} x'^{j}$, we obtain $\xi_{0}=1$,  $\xi_{1}=-1$ and $\cot(\omega'+\theta)=\tan\theta/\tan^{2}\varphi$. This means that the axis $x'^{0}$ can lie in the time-like region and the axis $x'^{1}$ lies in the space-like region. Making use of these conditions and Eqs.~(\ref{15})--(\ref{17}), we get
\begin{eqnarray}\label{18}
 a^{i}_{\ j} &=& \left(
               \begin{array}{cccc}
                 \gamma_{f}                                  & \gamma_{f}\tan\theta & 0 & 0 \\
                 \gamma_{f}\frac{\tan\theta}{\tan^{2}\varphi}  & \gamma_{f}          & 0 & 0 \\
                          0                                  &      0              & 1 & 0 \\
                          0                                  &      0              & 0 & 1 \\
               \end{array}
             \right), \\ \label{19}
    \gamma_{f} &=& \frac{1}{\sqrt{1-\frac{\tan^{2}\theta}{\tan^{2}\varphi}}}.
\end{eqnarray}
The geometrical illustration of the (generalized) Lorentz transformations (\ref{18}), (\ref{19}) can be seen in Fig.~\ref{fig.2}. The angle $\theta$ is restricted by the conditions $-\varphi \leq \theta\leq \varphi$ and $ \pi-\varphi \leq \theta\leq \pi+\varphi$ while the angle  $\omega'+\theta$ between the axes $x'^{0}$ and $x^{1}$ is restricted by the condition $\pi-\varphi \geq \omega'+\theta \geq \varphi$. For example, for $\theta=0$ and $\theta=\pi$, we have $\omega'+\theta=\pi/2$, i.e. at the inversion of the space-like axis $x'^{1}$ the inversion of the time-like axis $x'^{0}$ is forbidden. Thus, the inversion of time (energy) axis under the (generalized) Lorentz transformations in the GSTR and STR is not allowed by the invariance of the metric tensor. This possibility is illustrated in Fig.~\ref{fig.3}.

In {\it the coordinate space}, the generalized Lorentz transformations are explicitly obtained from the linear transformations $x^{i} = a^{i}_{\ j} x'^{j}$ with matrix (\ref{18}) on conditions that $x'=0$ in both time-like and light-like regions ($\vec{s}^{2}\geq 0$, $\xi=1$ and $\xi=0$) and $t'=0$ in the space-like region ($\vec{s}^{2}<0$, $\xi=-1$). Then we have
\begin{eqnarray}\label{20}
  \tan \theta &=& \frac{v_{f}c}{v_{l}^{2}},  \qquad \xi=1, \quad \xi=0,  \\ \label{21}
  \tan \theta &=& \frac{c}{v_{f}},  \qquad \xi=-1,
\end{eqnarray}
where $v_{f}=x/t$ is the velocity of motion along the $x$-direction of the inertial reference frame $K'$ relative to the laboratory frame $K$. Substituting Eqs.~(\ref{20}) and (\ref{21}) into Eqs.~(\ref{18}), (\ref{19}), we obtain
\begin{equation}\label{22}
 a^{i}_{\ j}=\left(
               \begin{array}{cccc}
                 \gamma_{f}                                  & \gamma_{f} \frac{c v_{f}}{v_{l}^{2}} & 0 & 0 \\
                 \gamma_{f}\frac{v_{f}}{c}                   & \gamma_{f}          & 0 & 0 \\
                          0                                  &      0              & 1 & 0 \\
                          0                                  &      0              & 0 & 1 \\
               \end{array}
             \right),
             \quad \xi=1, \;\; \xi=0,
\end{equation}
and
\begin{equation}\label{23}
 a^{i}_{\ j}=\left(
               \begin{array}{cccc}
                 \gamma_{f}                                  & \gamma_{f}\frac{c}{v_{f}}   & 0 & 0 \\
                 \gamma_{f}\frac{v_{l}^{2}}{c v_{f}}                   & \gamma_{f}          & 0 & 0 \\
                          0                                  &      0              & 1 & 0 \\
                          0                                  &      0              & 0 & 1 \\
               \end{array}
             \right),
              \quad \xi=-1,
\end{equation}
where
\begin{eqnarray}\label{24}
  \gamma_{f}&=& \frac{1}{\sqrt{1-\frac{v_{f}^{2}}{v_{l}^{2}}}}, \qquad  \xi=1, \quad \xi=0, \\ \label{25}
  \gamma_{f}&=& \frac{1}{\sqrt{1-\frac{v_{l}^{2}}{v_{f}^{2}}}}, \qquad  \xi=-1.
\end{eqnarray}
For the non-degenerate transformations (\ref{22}), (\ref{23}) there is the inverse matrix $b^{i}_{\ j}$ which is determined from the relation~\cite{Pauli,Charlier}
\begin{equation}\label{26}
   b^{i}_{\ j}a^{j}_{\ k}= a^{i}_{\ j}b^{j}_{\ k} = \delta^{i}_{\ k}.
\end{equation}
The explicit form of the inverse matrix $b^{i}_{\ j}$ can be obtained from Eqs.~(\ref{22}), (\ref{23}) by changing $v_{f}$ to $-v_{f}$.

In particular, the generalized Lorentz transformations (\ref{22})--(\ref{25}) for the contravariant and covariant components, $x^{i} = a^{i}_{\ j} x'^{j}$ and  $x_{i} =  x'_{j} b^{j}_{\ i}$~\cite{Charlier}, in both the time-like and the light-like regions $(\vec{s}^{2}\geq 0)$~\cite{Cardone} and in the space-like region $(s^{2}< 0)$ can be written as
\begin{eqnarray}\label{27}
               t&=&\gamma_{f} \left(t'+ \frac{v_{f}}{v_{l}^{2}} x'\right), \qquad \xi=1, \;\; \xi=0, \\  \label{28}
               x&=&\gamma_{f} \left(v_{f} t'+ x'\right), \\  \label{29}
               y&=&y', \\  \label{30}
               z&=&z'
\end{eqnarray}
and
\begin{eqnarray}\label{31}
               t&=&\gamma_{f} \left(t'+ \frac{1}{v_{f}} x'\right), \qquad \xi=-1, \\  \label{32}
               x&=&\gamma_{f} \left(\frac{v_{l}^{2}}{v_{f}} t'+ x'\right), \\  \label{33}
               y&=&y', \\  \label{34}
               z&=&z'.
\end{eqnarray}
In {\it momentum space}, the generalized Lorentz transformations (\ref{22})--(\ref{25}) are
\begin{eqnarray}\label{35}
               E&=&\gamma_{f} \left(E'+ \frac{v_{f}c^{2}}{v_{l}^{2}} p_{x}'\right), \quad \xi=1, \;\; \xi=0, \\  \label{36}
               p_{x}&=&\gamma_{f} \left(\frac{v_{f}}{c^{2}} E'+ p_{x}'\right), \\  \label{37}
               p_{y}&=&p_{y}', \\  \label{38}
               p_{z}&=&p_{z}'
\end{eqnarray}
and
\begin{eqnarray}\label{39}
               E&=&\gamma_{f} \left(E'+ \frac{c^{2}}{v_{f}} p_{x}'\right), \quad \xi=-1, \\  \label{40}
               p_{x}&=&\gamma_{f} \left(\frac{v_{l}^{2}}{c^{2}v_{f}} E'+ p_{x}'\right), \\  \label{41}
               p_{y}&=&p_{y}', \\  \label{42}
               p_{z}&=&p_{z}'.
\end{eqnarray}
Using Eqs.~(\ref{27})--(\ref{34}), the generalized Lorentz transformations for the components of the $3$-velocity can be written as
\begin{eqnarray}\label{45}
    v_{x} &=& \gamma_{x}(v_{f}+v'_{x}), \qquad \xi=1, \;\; \xi=0,  \\ \label{46}
    v_{y} &=& \frac{\gamma_{x}}{\gamma_{f}}v'_{y}, \\ \label{47}
    v_{z} &=& \frac{\gamma_{x}}{\gamma_{f}}v'_{z}
\end{eqnarray}
and
\begin{eqnarray}\label{48}
    v_{x} &=& \gamma_{x}\left(\frac{v_{l}^{2}}{v_{f}}+v'_{x}\right), \qquad \xi=-1, \\ \label{49}
    v_{y} &=& \frac{\gamma_{x}}{\gamma_{f}}v'_{y}, \\ \label{50}
    v_{z} &=& \frac{\gamma_{x}}{\gamma_{f}}v'_{z},
\end{eqnarray}
where
\begin{eqnarray}\label{51}
\gamma_{x}&=&\frac{1}{1+\frac{v_{f}v'_{x}}{v_{l}^{2}}}, \qquad \xi=1, \;\; \xi=0, \\ \label{52}
\gamma_{x}&=&\frac{1}{1+\frac{v'_{x}}{v_{f}}}, \qquad \xi=-1,
\end{eqnarray}
with $\gamma_{f}$ given by Eqs.~(\ref{24}), (\ref{25}). The four-interval (\ref{7}) and four-momentum (\ref{8}) are invariants under the transformations (\ref{27})--(\ref{34}) and (\ref{35})--(\ref{42}), respectively,
\begin{eqnarray}\label{53}
    v_{l}^{2}t^{2}-x^{2}-y^{2}-z^{2} &=&  v_{l}^{2}t'^{2}-x'^{2}-y'^{2}-z'^{2},           \\ \label{54}
   \frac{v_{l}^{2}}{c^{4}}E^{2}-p_{x}^{2}-p_{y}^{2}-p_{z}^{2} &=& \frac{v_{l}^{2}}{c^{4}}E'^{2}-p_{x}'^{2}-p_{y}'^{2}-p_{z}'^{2}. \;\;\;\;\;
\end{eqnarray}
The transformations on the basis vectors and dual basis vectors are given by $\vec{e}_{i}=\vec{e}'_{j} b^{j}_{\ i}$ and $\vec{e}^{\ i} = a^{i}_{\ j} \vec{e}'^{j}$. The inverse transformations on the contravariant and covariant components are $x'^{i} = b^{i}_{\ j} x^{j}$ and $x'_{i} =  x_{j} a^{j}_{\ i}$ and on the basis vectors and dual basis vectors are $\vec{e}'_{i}=\vec{e}_{j} a^{j}_{\ i}$ and $\vec{e}'^{\ i} = b^{i}_{\ j} \vec{e}^{j}$~\cite{Charlier}. The explicit inverse formulas for the generalized Lorentz transformations are most easily obtained from Eqs.~(\ref{27})--(\ref{34}) and Eqs.~(\ref{35})--(\ref{42}) by exchanging the primed and unprimed coordinates and changing $v_{f}$ to $-v_{f}$.

\begin{figure}
\onefigure[width=8.5cm]{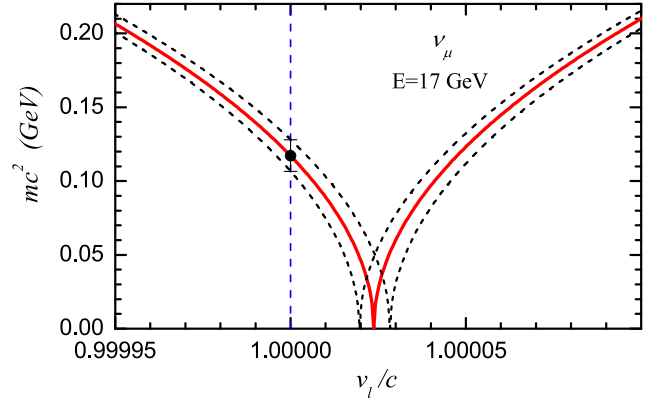}
\caption{(Color online) The dependence of the muon neutrino mass $m$ on the speed limit $v_{l}$ at fixed energy $E=17$ GeV and velocity $(v-c)/c=2.37^{+0.47}_{-0.40}\times 10^{-5}$ for the OPERA neutrino experiment~\cite{Opera}. The dotted lines represent the combined statistical and systematic errors. The symbol represents the mass of the STR superluminal space-like neutrino.}
\label{fig.4}
\end{figure}

\section{Results and discussions}
It is pertinent to mention that the STR is a particular case of the GSTR. If we assume $\varphi=\pi/4$, which resembles the equilateral hyperboloid of revolution, $a=b$, and provides an equivalence of the limiting velocity to the speed of light, $v_{l}=c$, then all relations of the STR and the Lorentz transformations are recovered.

Figure~\ref{fig.4} presents the behavior of the mass $m$ of the muon neutrino $\nu_{\mu}$ as a function of its speed limit $v_{l}$ which was calculated by Eqs.~(\ref{12}), (\ref{14}) at the fixed neutrino energy $E$ and velocity $v$ for the OPERA experiment~\cite{Opera}. The deviation of the neutrino velocity from its speed limit results in a significant increase of the neutrino mass. If the speed limit of the OPERA neutrino with the speed $v$ is $v_{l}=c$, then the neutrino is to be a STR space-like superluminal particle with the mass $m=117.1^{+11.0}_{-10.5}$ MeV$/c^{2}$. Otherwise, for the small neutrino mass, $m=2$ eV$/c^{2}$~\cite{Wein}, the relative deviation from the limiting velocity $v_{l}$ of the neutrino velocity $v$ due to its finite rest mass is expected to be smaller than $|v_{l}-v|/v= 6.92 \times 10^{-21}$ and it is either a time-like or a space-like superluminal particle with the velocity of motion close to the limiting speed. If the neutrino velocity is equal to its speed limit, then its mass must be $m=0$ and it is a light-like superluminal particle. Thus, we conclude that the OPERA neutrino is a time-like or a light-like superluminal particle with the speed of propagation very close to the limiting velocity which exceeds the speed of light. To ensure that the OPERA neutrino is moving faster than light, it is necessary not only to measure its position at two times but also to measure its energy and momentum and calculate the velocity, $v_{x}=c^{2}p_{x}/E$~\cite{Feinberg}. At $E=17$ GeV, the momentum of the OPERA neutrino~\cite{Opera} should be $p_{x}=17.0004\pm 0.0001$ GeV$/c$.

\section{Conclusions}
To conclude, in this Letter we have deduced the GSTR with an arbitrary limiting velocity from the canonical equation of the $4$-dimensional hyperboloid of revolution and have investigated the possibility of describing particles that travel faster than light. We have found the generalized Lorentz transformations for any values of the speed limit for both the time-like and the spice-like regions in the coordinate and momentum spaces and in the general parametric form. It was obtained that the metrics and the Lorentz transformations of the STR correspond to the one particular case of the metrics and generalized  Lorentz transformations of the GSTR, when the hyperboloid of revolution resembles the equilateral hyperboloid of revolution at $\varphi=\pi/4$ which corresponds to the speed limit equal to the light velocity. We have found that in the space-like region the rest mass of a particle is real due to the unified definition of the square length of the four-vector from the canonical equation of the hyperboloid of revolution. It was obtained that the (generalized) Lorentz transformations for a four-interval, a four-momentum and a $3$-velocity in the space-like and the time-like regions are different; however, the metrics of the space-time in these regions are the same. We have found that the inversion of time (energy) axis under the (generalized) Lorentz transformations in the GSTR and STR is forbidden by the invariance of the metric tensor. The spacial subspace of the four-dimensional Minkowski space is complex. In the GSTR, any value of the limiting velocity is invariant in all inertial reference frames likewise the speed of light. Therefore, the GSTR deals with three distinct classes of superluminal particles: time-like, space-like and light-like classes. We have revealed that the muon neutrino in the OPERA experiment is most likely a light-like or a time-like superluminal particle rather than a space-like superluminal particle. Thus, this experiment suggests that the limiting velocity of the muon neutrino may be greater than the velocity of light in vacuum.

\acknowledgments
This work was supported in part by the joint research project of JINR and IFIN-HH, protocol N~4063.

\end{document}